\documentclass[apj]{emulateapj}

\usepackage{apjfonts,graphicx,subfigure,epsfig,natbib}

\def\starlight{\textsc{starlight}}
\newcommand\ebv{\text{$E(B\!-\!V)$}}
\newcommand\esbv{\text{$E_s(B\!-\!V)$}}
\newcommand\egbv{\text{$E_g(B\!-\!V)$}}

\shorttitle{Study the ULIRGs Evolutionary Scenario through IRAS\,F13308+5946}
\shortauthors{Meng et al.}
\journalinfo{Accepted for publication in ApJ}
\submitted{Accepted for publication in ApJ}

\begin{document}

\title{IRAS\,F13308+5946: A Possible Transition Phase From Type I ULIRG To Optical Quasar}

\author{Xian-Min Meng\altaffilmark{1,2}, Hong Wu\altaffilmark{1}, Qiu-Sheng Gu\altaffilmark{3}, Jing Wang\altaffilmark{1}, Chen Cao\altaffilmark{4}}
\altaffiltext{1}{National Astronomical Observatories, Chinese Academy of Sciences, Beijing
             100012, China; {\it mengxm@nao.cas.cn}}
\altaffiltext{2}{Graduate University of Chinese Academy of Sciences, Beijing 100039, China}
\altaffiltext{3}{Department of Astronomy, Nanjing University, Nanjing 210093, China}
\altaffiltext{4}{Shandong University at Weihai, Weihai 264209, China}


\begin{abstract}
We present a stellar population synthesis study of a type I luminous infrared galaxy (LIRG): IRAS\,F13308+5946.
It is a quasar with absolute magnitude $M_i=-22.56$ and has a spectral feature of a Seyfert 1.5 galaxy.
Optical images show characteristics of later stages of a merger.
With the help of the stellar synthesis code \starlight\ (Cid Fernandes et al.\ 2005)
and both Calzetti et al.\ (2000) and Leitherer et al.'s (2002) extinction curves,
we estimate the past infrared (IR) luminosities of the host galaxy 
and find it may have experienced an ultraluminous infrared galaxy (ULIRG) phase for nearly 300\,Myr,
so this galaxy has probably experienced a type I ULIRG phase.
Both nuclear starburst and active galactic nuclei (AGN) contribute to the present IR luminosity budget,
and starburst contributes $\sim$70\%. 
The mass of supermassive black-hole (SMBH) is $M_{\rm BH}=1.8\times10^8\,M_\odot$
and the Eddington ratio $L_{\rm bol}/L_{\rm Edd}$ is 0.12, which both approximate to typical values of PG QSOs.
These results indicate that IRAS\,F13308+5946 is probably at the transitional phase from a type I ULIRG to a classical QSO.
\end{abstract}


\section{Introduction}

Luminous infrared galaxies (LIRGs) and ultraluminous infrared galaxies (ULIRGs) are galaxies
with infrared (IR) luminosities $L_{\rm IR} = L(8\text{--}1000\mu{\rm m}) > 10^{11}$\,$L_{\odot}$ and
$L_{\rm IR} > 10^{12}$\,$L_{\odot}$, respectively.
Their IR luminosities and space densities are both comparable to
QSOs (Sanders \& Mirabel 1996). These galaxies always show signs of tidal interaction and merger, 
and the interaction rate is observed to increase with IR luminosity
(Zou et al.\ 1991; Clements et al.\ 1996; Murphy et al.\ 1996; Wu et al.\ 1998; Veilleux et al.\ 2002).
Nearly all ULIRGs are strong interaction or merger systems (Kim et al.\ 2002).

It is widely accepted that galaxy interactions and mergers trigger extreme nuclear activity, 
as well as more widespread starburst (Toomre \& Toomre 1972; Larson \& Tinsley 1978). 
The molecular gas concentrations of ULIRGs are in their central kpc regions 
(Downes \& Solomon 1998) and have the ability to form stars with densities comparable to those 
in elliptical galaxies (Tacconi et al.\ 2002). Kormendy \& Sanders (1992) proposed that 
ULIRGs may evolve into elliptical galaxies through merger induced, dissipative collapse. 
Structural, kinematic and photometric properties of ULIRGs show that they are originated 
from major mergers and they averagely fall on the fundamental plane of moderate-mass ellipticals 
(stellar mass $\sim 10^{10}$--$10^{11}\,M_\odot$), but are well offset from giant ellipticals, 
suggesting that ULIRG mergers are ellipticals in formation 
(Genzel et al.\ 2001; Tacconi et al.\ 2002; Dasyra et al.\ 2006a). 

Observations revealed that a large fraction of LIRGs/ULIRGs show spectral characteristics 
as Seyfert galaxies (Wu et al.\ 1998; Kewley et al.\ 2001). 
The fraction increases dramatically with IR luminosity, and among ULIRGs, the fraction is about 50\% 
(Kim et al.\ 1998a; Veilleux et al.\ 1999; Cao et al.\ 2006; Yuan et al.\ 2010; Nardini et al.\ 2010), 
or even higher ($\sim$70\%; Nardini et al.\ 2010). 
Though many of these sources are dominated by AGN in their bolometric luminosity 
(Boller et al.\ 2002; Nandra \& Iwasawa 2007), in most cases the predominance of starburst 
over AGN is proposed (Gu et al.\ 1997; Lutz et al.\ 1999; Franceschini et al.\ 2003), 
and Nardini et al.\ (2008) proposed a fraction of $\sim85$\%.

From millimeter-wave CO observations and optical spectra analyse of 10 ULIRGs, 
Sanders et al.\ (1988) proposed that ULIRGs represent the dust-enshrouded stage of QSOs. 
They proposed a classical evolution scenario that two gas-rich spirals merge 
(or interact) first and the funneled gas toward the merger center triggers nuclear starburst 
before the ignition of a dust-enshrouded AGN. 
When dust has been consumed or swept away under the radiative pressure of AGN and supernovae,
an optical quasar would appear.
Canalizo \& Stockton (2000, 2001) selected a sample of 9 low-redshift QSOs which fall onto the
intermediate position between the regions occupied by ULIRGs and QSOs in the far-infrared (FIR)
color-color diagram. All these 9 transition QSOs are
undergoing tidal interactions and 8 are major mergers.
All of them also show strong recent star formation activity within 300\,Myr.
They proposed a model involving a dust cocoon that initially surrounds the QSO nuclear regions,
which can account for all the observed and derived properties in transition objects, 
and it is consistent with the idea of Sanders et al.\ (1988). 
They suggested that either at least some ULIRGs evolve to become classical QSOs,
or some QSOs are born under the conditions as ULIRGs and their lifetimes as QSOs
last $\leq $ 300\,Myr.
Most of these transition QSOs are also ULIRGs. As Kawakatu et al.\ (2006), we refer to
QSOs and Seyfert 1 galaxies that selected from ULIRGs as type I ULIRGs
(or IR QSOs in Zheng et al.\ 2002). Kawakatu et al.\ (2006) and Hou et al.\ (2009) find that
type I ULIRGs are the early phase of black-hole (BH) growth and they are QSOs in formation.

The case may be as Colina et al.\ (2001) had proposed that high-luminosity QSOs would be the end
point in the merging process of massive ($>L$*) disk galaxies, while cool ULIRGs, which would be
the end product in the merging of two or more low-mass (0.3$L$*--0.5$L$*) disk galaxies, would
not evolve into QSOs. This scenario is supported by many studies (McLeod \& Rieke 1994;
Arribas et al.\ 2003; Dasyra et al.\ 2006a).

In this paper we present the result of stellar population synthesis of IRAS\,F13308+5946 and
estimations of past IR luminosities. IRAS\,F13308+5946 is a LIRG at present and
shows a spectral characteristic of a Seyfert 1.5 galaxy,
which together suggest a potential transition phase between type I ULIRGs and classical QSOs.
It shows clear absorption lines and young stellar population features for spectral fitting.
Observations from {\it IRAS}, 2MASS, SDSS and
FIRST are presented in \S2. We describe the process of extracting the star formation
history (SFH), emission-line fitting, aperture correction and the extinction curve in \S3.
In \S4 we present some estimations like the current star formation rate (SFR), the
BH mass, the Eddington ratio, etc., based on the emission-lines fitting.
The past IR luminosities are estimated in \S5. Conclusions and discussions are given in \S6.
We summarize in \S7.
Through out this paper we adopt a cosmology of $H_0$=71 km s$^{-1}$ Mpc$^{-1}$, 
$\Omega_M$=0.29 and $\Omega_{\Lambda}$=0.71.

\section{Observations}

IRAS\,F13308+5946 was observed by the Infrared Astronomical Satellite ({\it IRAS}), 
Two Micron All Sky Survey (2MASS) All-Sky Extended Source Survey 
(recorded as 2MASX J13323783+5930538),
Sloan Digital Sky Survey (SDSS, recorded as SDSS J133237.94+593053.7) and 
Very Large Array (VLA) Faint Images of the Radio Sky at Twenty-centimeters
(FIRST) survey. It hasn't been explored by UV or X-ray observations by far.

\subsection{Mid- and Far-Infrared}
The mid-to-far IR fluxes of IRAS\,F13308+5946 are taken from the {\it IRAS} Faint Source
Catalog (FSC) Version 2.0. Flux densities at three wavebands, 12, 25 and 100 $\mu$m, are upper limits. 
Only the one at 60 $\mu$m, which is 0.2615 Jy, has a record with high quality (FQUAL=3). 
Therefore, the IR luminosity ($L_{\rm IR}(8\text{--}1000 \mu{\rm m})$) is calculated by
\begin{equation}
\label{Lir1} L_{\rm IR}(8\text{--}1000 \mu{\rm m}) \approx 2 L_{60 \mu{\rm m}},
\end{equation}
(Lawrence et al.\ 1989; Bushouse et al.\ 2002; Arribas et al.\ 2004; Wang et al.\ 2008), 
where $L_{\rm 60 \mu{\rm m}}$ is the luminosity at 60 $\mu$m. 
The derived IR luminosity is $L_{\rm IR}(8\text{--}1000 \mu{\rm m}) = 10^{11.56}\,L_{\odot}$
$\footnote{The luminosity in the wavelength range 1--8 $\mu$m is estimated to be
less than $\approx$5\%--10\% of the total and negligible (Calzetti et al.\ 2000).}$.
For comparison, we also calculate IR luminosity using an IR spectral energy distribution (SED) model. 
Since IRAS\,F13308+5946 is not only a starburst(SB) galaxy, it also has a type 1.5 AGN (see \S3.4), 
we adopt the SED model established by Siebenmorgen et al.\ (2004) for galaxies with both AGN and starburst components. 
After the SED model was scaled to the observed 60$\mu$m flux, the integrated IR luminosity is calculated to be 
$L_{\rm IR}(8\text{--}1000 \mu m) = 10^{11.55}\,L_{\odot}$, exactly the same as the one above.

We also calculate the IR luminosity in the range 1--1000 $\mu$m through 60 $\mu$m and 100 $\mu$m fluxes 
(Lonsdale-Persson \& Helou 1987; Calzetti et al.\ 2000):
\begin{equation}
\label{Ffir}
F_{\rm FIR} (40\text{--}120 \mu{\rm m}) = 1.26 \times 10^{-14} \lbrace2.58 f_{60 \mu{\rm m}} + f_{100 \mu{\rm m}}\rbrace [{\rm W~m}^{-2}]
\end{equation}
\begin{equation}
\label{fir2}
F_{\rm IR}(1\text{--}1000 \mu{\rm m}) = (1.75 \pm 0.25) F_{\rm FIR} (40\text{--}120 \mu{\rm m})
\end{equation}
\begin{equation}
\label{Lir2}
L_{\rm IR}(1\text{--}1000 \mu{\rm m}) = 4 \pi D_{\rm L}^{2} F_{\rm IR}(1\text{--}1000 \mu{\rm m}) L_{\rm \odot}.
\end{equation}
The derived IR luminosity $L_{\rm IR}(1\text{--}1000 \mu{\rm m}) = 10^{11.8 \pm 0.07}\,L_{\rm \odot}$ 
can be used as upper limit, and it is consistent with the result only adopting $f_{60\mu{\rm m}}$. 
Therefore, we adopt $L_{\rm IR}(8\text{--}1000 \mu{\rm m})
= 10^{11.56}\,L_{\rm \odot}$ as the IR luminosity throughout this
paper. We don't do {\it K}-corrections on IR fluxes, because the
correction factor is small ($\sim$5\% adopting SB SED model from Siebenmorgen et al. 2007) 
enough to neglect, compared with the uncertainty of IR luminosity calculation and 
the scatter among different {\it K}-correction methods.

\subsection{Optical and Near-infrared}
IRAS\,F13308+5946 was also observed by SDSS imaging camera and
spectrograph. The {\it u-, g-, r-, i-, z}-band images and the spectrum
are from the SDSS Data Release 5 (DR5). Each band was observed with 
exposure time of 53.9\,s, and the pixel size is 0.396\arcsec\ on the sky. 
The spectrum has been corrected for Galactic reddening of 
$\ebv=0.015$ and redshift of $z=0.171$, so the rest-frame wavelength coverage 
is 3258--7852\AA.

The SDSS spectrum shows a typical feature of a type I AGN. 
The {\it i}-band absolute magnitude of $M_i = -22.564$ satisfies the quasar criterion of $M_i < -22.0$ 
defined by Schneider et al.\ (2007).
However, the {\it B}-band absolute magnitude is $M_B = -21.1$ 
(transferred from $M_B = -21.8$ which adopts $H_0 = 50$ km s$^{-1}$ Mpc$^{-1}$, V\'eron-Cetty \& V\'eron 2006), 
and it doesn't satisfy the criterion for quasars of $M_B< -22.2$ ($H_0=71$ km s$^{-1}$ Mpc$^{-1}$) 
defined by Schmidt \& Green (1983). 
The larger extinction at {\it B}-band could explain the inconsistency between the two band (see also \S3.6 ).

\begin{figure}[htb]
\begin{minipage}[t]{0.45\textwidth}
\centering
\includegraphics[width=72mm]{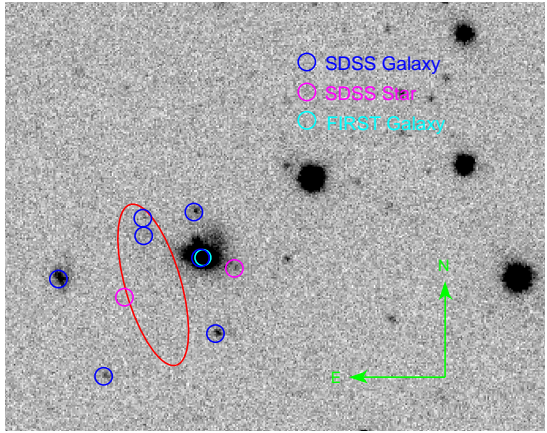}
\caption{Source identification from {\it IRAS} and SDSS. The red ellipse is {\it IRAS}
uncertainty ellipse with 2.45-$\sigma$ confidence for each axis, which is plotted on
SDSS {\it r}-band image. {\it Blue} circles represent SDSS galaxies and {\it magenta} ones SDSS stars.
The {\it cyan} circle marks the FIRST source in the field of view, 
and coincides with the galaxy IRAS\,F13308+5946.}
\label{f1eps}
\end{minipage}
\end{figure}

There is obvious deviation between SDSS and {\it IRAS}
positions. The {\it IRAS} 2.45-$\sigma$ uncertainty ellipse 
(corresponding to 95\% probability enclosure) with semi-major and 
semi-minor axises of 21\arcsec\ and 7\arcsec\ does not
embrace the optical (SDSS) position (see Fig.~\ref{f1eps}). The optical
source only locates within the 5-$\sigma$ uncertainty ellipse of {\it IRAS} position. 
Since there is not any other bright sources within {\it IRAS} uncertainty ellipse in either
optical and near-infrared (2MASS) bands, and IR phenomenons are most 
related with galactic activities, we identify the object 
detected by {\it IRAS}, 2MASS and SDSS as the same one. In \S2.3,
the well-known radio-FIR correlation will confirm such 
identification. Hereafter, we refer to the object as IRAS\,F13308+5946.

\begin{figure*}[htb]
\centering
\includegraphics[width=150mm]{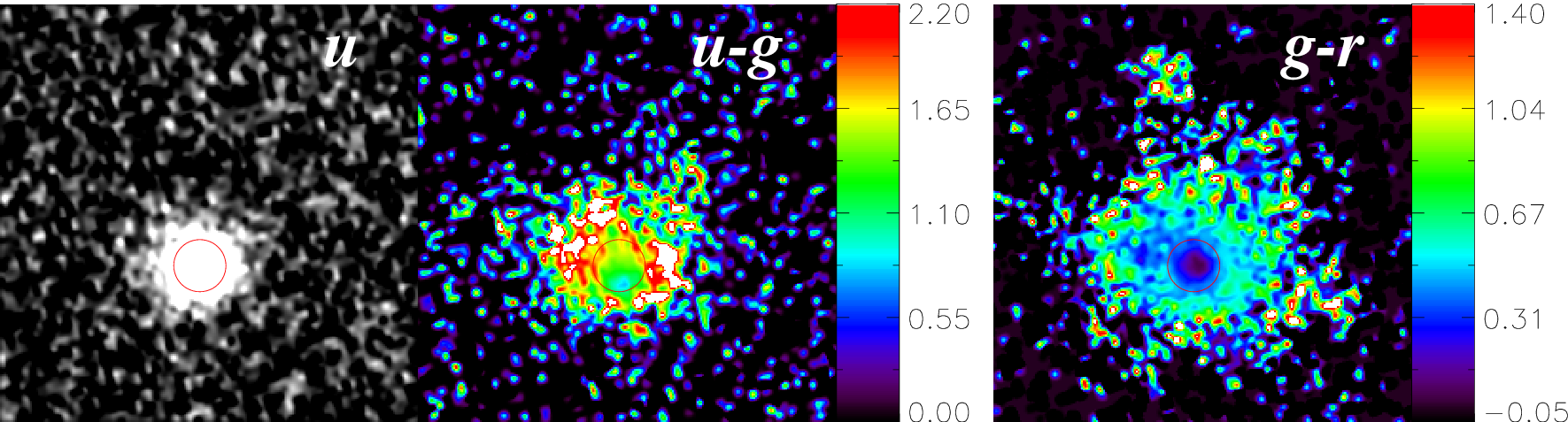}
\caption{SDSS {\it u}-band image and $u\!-\!g$, $g\!-\!r$ color maps of IRAS\,F13308+5946.
The {\it red} circle represents coverage of the 3\arcsec\ diameter fiber aperture on the host galaxy. 
The FOV of the three images are all 25\arcsec$\times$25\arcsec.}
\label{f2eps}
\end{figure*}
Fig.~\ref{f2eps} ({\it left}) gives the {\it u}-band image and the
3\arcsec\ diameter fiber of the spectrograph, which covers
$\sim$10 kpc of the galaxy. Two color-map images of {\it $u\!-\!g$} and {\it $g\!-\!r$} are also presented. 
The field of view (FOV) of these images are all 25\arcsec$\times$25\arcsec. These images are
made after world coordinate system matching, Galactic reddening correction 
and {\it K}-correction (IDL routine from Blanton et al.\ 2003, v4\_1\_4). 
Because the seeings at {\it u-, g-, r}-bands are 1.3\arcsec, 
1.3\arcsec\ and 1.1\arcsec\ respectively, we have to match the seeings by convolving the {\it r}-band 
image with a 2-D Gaussian kernel with $\sigma=0.29$\arcsec\ before making $g\!-\!r$ image. 
The $g\!-\!r$ image shows a color gradient that the central 
region of galaxy is quite blue. It could be explained by 
circumnuclear starburst and/or AGN activity. After converting 
SDSS $g\!-\!r$ color to \bv\ color (Jester et al.\ 2005), we 
obtained the \bv\ color varies from $\sim$0.25 at the center, 
to $\sim$0.5 at the edge of the fiber aperture and to $>$1 at the 
outskirt of the galaxy. If these variations are due to different 
stellar populations, the ages of the stellar populations may range 
from $\sim$300\,Myr to $\sim$800 (assuming solar metallicity) 
inside the aperture, and to $>$10\,Gyr at the galaxy outskirt, 
adopting the stellar populations evolution model of Bruzual \& Charlot (2003; hereafter BC03). 
Considering local dust extinction, the $g\!-\!r$ color may be reddened, so the ages 
obtained above should be somewhat younger. Generally, the $u\!-\!g$ 
color distribution is consistent with $g\!-\!r$ color according to BC03. 
However, unlike $g\!-\!r$ color, the bluest region on the $u\!-\!g$ image deviates 
from the galaxy center. It could be attributed to the merger process. 

From Fig.~\ref{f1eps} and Fig.~\ref{f2eps} we can find characteristics of later stages of a merger. 
Adopting 2MASS standard isophotal photometry, the {\it H}-band absolute magnitude is 
$M_H=-25.7$, which is 4 times as luminous as $L$* galaxies ($M_H=-24.2$; Colina et al.\ 2001). 
Therefore, the galaxy could be a merger system of massive ($\geq L$*) galaxies.

\subsection{Radio}
The only one FIRST source in the FOV of Fig.~\ref{f1eps} is marked with a {\it cyan} circle. 
The radio source coincides with the optical quasar with a spacial separation of 0.63\arcsec. 
The 1.4 GHz continuum integrated flux density detected by FIRST is 0.93 mJy. 
It follows the well-known radio-FIR correlation which is confirmed by 
Yun et al.\ (2001) by the overwhelming majority ($\geq$98\%) of 1809 IR-selected galaxies 
($S_{60 \mu{\rm m}}\geq2\,{\rm Jy}$). 
Moreover, the derived parameter $\log R^*=\log{f(5000\,\text{MHz}) \over f(2500\,\text{\AA})}=0.423$ 
(Sramek \& Weedman 1980) is much lower than $1$, indicating that it belongs to radio-quiet populations 
(Stocke et al.\ 1992). 
The coincidence between optical and radio position and its radio-FIR relation support that 
the optical and the IR source is the same one. 

\section{Spectral Analysis}
\subsection{Spectral Synthesis with \starlight}
We use the spectral synthesis code \starlight\ (Cid Fernandes et al.\ 2005) 
to derive stellar populations of the host galaxy. \starlight\ fits the observed spectrum $O_\lambda$ with a 
model spectrum $M_\lambda$ which is made up of a pre-defined set of base spectra. 
There is no need to give the value of parameters initial guess. 
It carries out the fitting with a simulated annealing plus Metropolis scheme 
to yield the minimum $\chi^2 = \sum_\lambda [(O_\lambda - M_\lambda) w_\lambda]^2$, 
where $w_\lambda^{-1}$ is the error in $O_\lambda$ at each wavelength. 
It models $M_\lambda$ by a combination that
\begin{equation}
\label{Mlambda}
M_\lambda = M_\lambda(\vec{x}, A_V, v_\star, \sigma_\star) =
\sum_{j=1}^{N_\star} x_j \gamma_{j,\lambda} r_\lambda
\end{equation}
where $\gamma_{j,\lambda} \equiv b_{\lambda,j} \otimes G(v_\star, \sigma_\star)$,
$b_{\lambda,j} \equiv {B_{\lambda,j}\overwithdelims () B_{\lambda_{0},j}}$
is the normalized flux of the $j^{th}$ spectrum,
$B_{\lambda,j}$ is the $j^{th}$ component of base spectrum,
$B_{\lambda_{0},j}$ is the value of the $j^{th}$ base spectrum at the normalization
wavelength $\lambda_0$, $G(v_\star, \sigma_\star)$ is the Gaussian distribution 
centered at velocity $v_\star$ and with dispersion $\sigma_\star$ of the
line-of-sight stellar velocity, $x_j$ is the fraction of light due to component {\it j}
at $\lambda_0$, $r_\lambda \equiv 10^{-0.4(A_\lambda-A_V)}$ is the global
extinction term and represented by $A_V$. 
The nebular and AGN emission lines can be obtained by subtracting the model spectrum from
the observed one as $E_\lambda = O_\lambda - M_\lambda$. 

\starlight\ uses random Markov Chains which needs to be appointed an integer seed to generate random numbers 
during the fitting, and if it uses different seeds, it will give different results 
even if the parameters are configured identically. Such difference will not change
the overall populations distribution significantly, but individual component $x_j$ may vary
$\sim$10 percentage in some cases. To obtain a statistically reliable result,
we carried out fittings with a set of seeds generated 
by Monte Carlo sampling and adopt the mean value over all seeds for every parameter.

\subsection{Parameters Determination}
We take simple stellar populations (SSPs) from BC03 and a power-law spectrum of AGN 
(if needed) $F_{\rm \lambda} \propto \lambda^{\alpha_{\rm \lambda}}$
as our base spectra. To minimize the parameter space and to make the
result more reliable, we adopt spectral templates with 10 ages
(0.005, 0.025, 0.1, 0.29, 0.5, 0.9, 1.4, 2.5, 4 and 10\,Gyr) computed with
``Padova 1994" evolutionary tracks (Alongi et al.\ 1993; Bressan et al.\ 1993;
Fagotto et al.\ 1994a,b; Girardi et al.\ 1996) and Chabrier (2003) initial mass function (IMF).

As a type I galaxy, the spectrum of IRAS\,F13308+5946 is possible to contain
a power-law component, though it is not obvious in the observed spectrum.
Since \starlight\ can include a power-law spectrum in the template base, we can carry out test
fittings to confirm whether it exists and find out the value of spectral index $\alpha_{\rm \lambda}$. 
A widely adopted AGN power-law slope ($F_{\rm \nu} \propto \nu ^ {\alpha_{\rm \nu}}$) over the optical
and UV region is $\alpha_{\rm \nu} = -0.5$ (Richstone \& Schmidt 1980),
while other studies had given the values from $-$1 to 0 (Natali et al.\ 1998, and
references therein), corresponding to $-2 \leq \alpha_{\rm \lambda} \leq -1$.
We carry out test fittings with $\alpha_{\rm \lambda}$ varies from $-$3.0 to $-$0.1
at intervals of 0.1 to search for the power-law index, and additional non-power-law fittings with
$\alpha_{\rm \lambda}=0$.

We optimize the fitting with single metallicity SSPs. To search for the best-fit metallicity, 
we carry out metallicity test with the six metallicities (Z = 0.0001, 0.0004, 0.004, 0.008, 0.02, 0.05)
of BC03 model for every power-law index. After that, we adopt the best-fit metallicity 
and power-law index for the formal fitting.
\begin{figure}[h]
\includegraphics[width = 90mm]{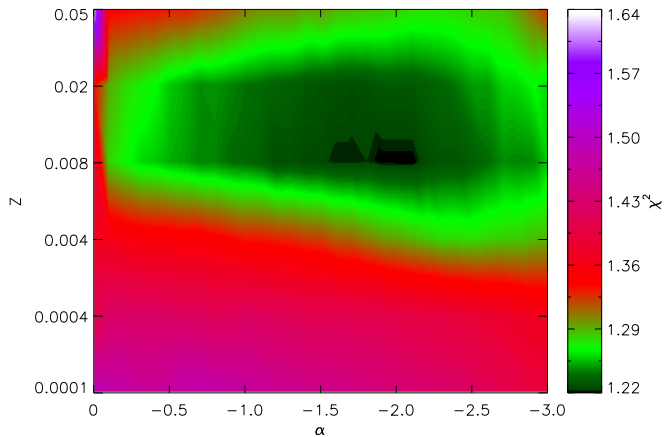}
\caption{Power-law index and metallicity test. 
Every test fitting adopts one of the six metallicities, a certain power-law index and 10 SSPs, 
and is evaluated by averaged $\chi^2$s over 25 individual fittings with different seeds. 
The {\it left} panel shows the averaged $\chi^2$s versus metallicities (Z) and power-law indices ($\alpha$). 
The {\it right} panel gives the color code of $\chi^2$.}
\label{f3eps}
\end{figure}

\begin{figure}
\includegraphics[width = 84mm]{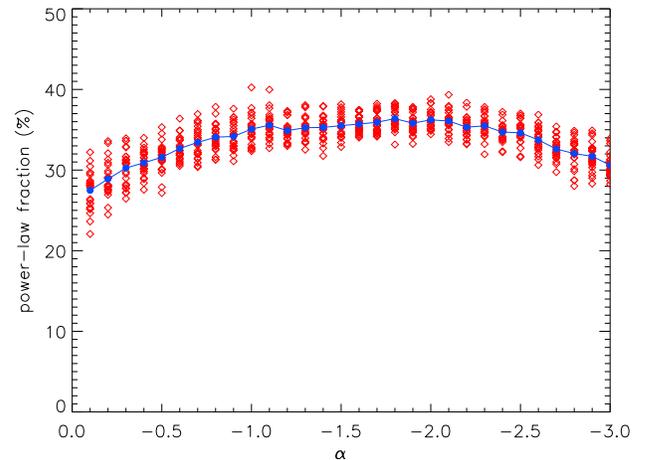}
\caption{AGN fractions in test fittings among different power-law indices. 
The {\it red unfilled diamonds} with a fixed power-law index represent data points of 25 different seeds. 
The {\it blue filled circles} represent the average AGN fraction of the 25 points.}
\label{f4eps}
\end{figure}

Fig.~\ref{f3eps} shows the test fitting results with different power-law indices $\alpha_{\lambda}$ 
and the six metallicities. 
We evaluate the fitting quality by the minimum $\chi^2$, which is averaged over 25 fittings 
with different seeds (randomly and uniformly distributed between $-$100000 and 100000). 
In Fig.~\ref{f3eps}, the minimum $\chi^2$ set reaches its bottom at Z=0.008 and $\alpha_{\lambda} = -2.0$. 
This is consistent with other studies that metallicities near solar value prevail in 
starburst galaxies (Tadhunter et al., 2005; Pellerin \& Robert 2007), and also that the 
power-law continuum slope falls around $\alpha_{\lambda} = -2$ 
(Francis 1996; Vanden Berk et al.\ 2001; Letawe et al.\ 2007). 
Moreover, the power-law index doesn't affect the synthesis very much. 
We learn from Fig.~\ref{f3eps} that if we fix the metallicity, 
the variation of $\chi^2$ among different power-law indices is not significant. 
Its influence on power-law component fraction is showed in Fig.~\ref{f4eps}. 
These data points are all from the test fits above, but only with subsolar metallicity ($Z = 0.4Z_\odot$). 
The {\it red unfilled diamonds} with a fixed power-law index represent the AGN component fraction at 
4020\AA\ of 25 different integer seeds. 
We can see from the figure that in the normal range of power-law index in literature, 
say, $-2 \leq \alpha_\lambda \leq -1$, 
the average AGN fraction keeps almost constant, 
and the dispersion of the average AGN fraction is far less than the one from different seeds. 
This result indicates that the form of AGN component doesn't affect the synthesis very much, 
so neither it affect the stellar component significantly.

Therefore, we take the parameters $\alpha_{\lambda} = -2.0$, $Z = 0.008$, and 10 SSP ages
for formal fitting.

\subsection{Formal Fitting and Stellar Populations}
\begin{figure*}[t]
\centering
\includegraphics[width=160mm]{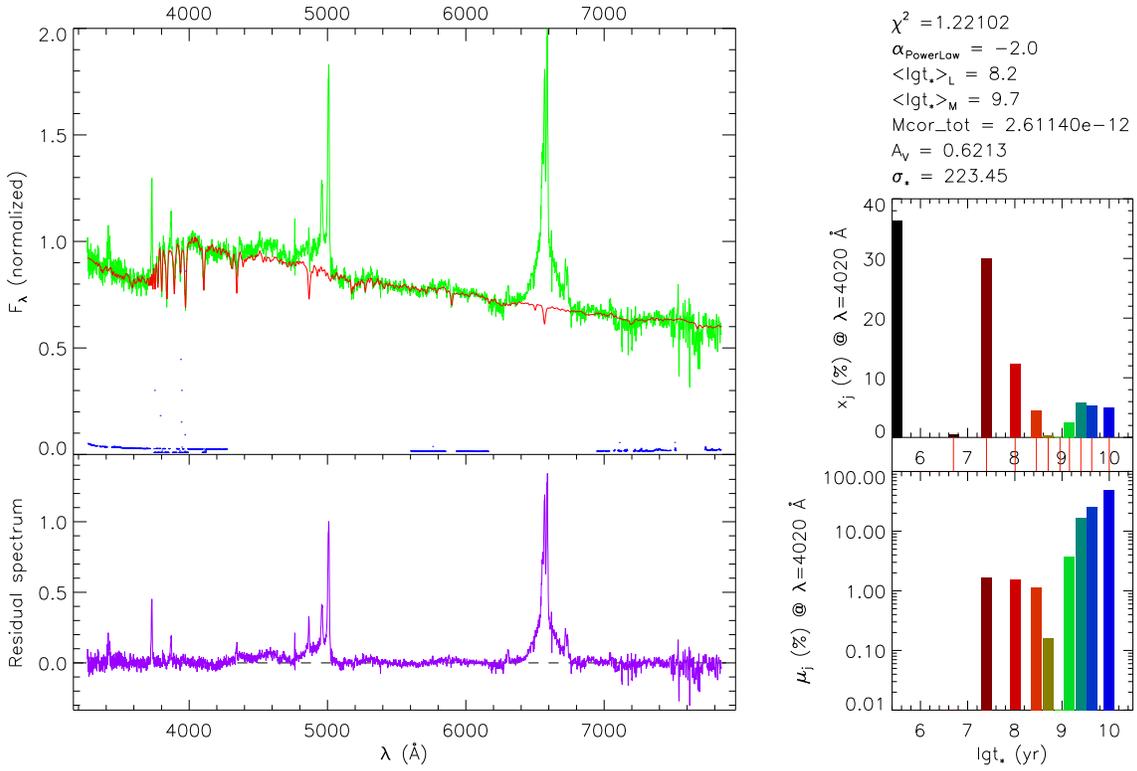}
\caption{{\small Spectral synthesis of IRAS\,F13308+5946. {\it Left-top}: The observed spectrum
$O_\lambda $({\it green}), the model spectrum $M_\lambda$ ({\it red})
and the error spectrum ({\it blue}) with the gaps meaning the masked regime
and the thrice weighted absorption lines.
{\it left-bottom}: The residual spectrum $E_\lambda$ ({\it purple}).
{\it Right}: Light ({\it top}) and mass ({\it bottom}) weighted stellar
population fractions $x_{\rm j}$ and $\mu_{\rm j}$, respectively.
The inserted panel on the right marks the ages of the stellar population templates.
The flux intensities of the {\it left} two panels are both normalized at 4020{\rm \AA} by
$4.549872\times10^{-16} \rm{(ergs\ s^{-1}\ cm^{-2})}$.}}
\label{f5eps}
\end{figure*}

The formal fitting involves 100 independent fittings with different seeds. 
Stellar absorption lines are given three times the weight larger than the continuum 
to emphasize the detail of stellar component. 
The recognized emission lines like [O $_{\rm II}$] $\lambda\lambda$3726+3729
([O $_{\rm II}$] $\lambda3727$),
Ne $_{\rm III}~\lambda$3869, H$\beta$+[O $_{\rm III}$] $\lambda\lambda$4959+5007,
H$\alpha$+[N $_{\rm II}$] $\lambda\lambda$6548+6583,
etc., together with NaI D5890 ISM absorption line are all masked before fitting.
The blended Fe $_{\rm II}$ emission-line regions are also
excluded, which are assigned to be 4270--4700 {\rm \AA} and 5100--5600 {\rm \AA}.
These masked regions are illustrated in Fig.~\ref{f5eps} represented by the gaps of the blue
curve on the {\it left-top} panel.
The blue curve is the error of the observed spectrum, and the inverse of error is the weight 
at each wavelength in spectral fitting. 
The observed spectrum $O_\lambda$ and all SSP templates are normalized at wavelength
$\lambda_0 = 4020$\AA\ before fitting.

The outcome fraction for each stellar component varies within $\sim$10 percentage point. 
Such variation is insignificant for main populations. 
We take the mean value for each parameter as the formal fitting result,
such as $x_{\rm j}$, $v_\star$, $\sigma_\star$, $\chi^2$, etc.
These parameters are all summarized in Fig.~\ref{f5eps}.
The {\it left-top} panel shows the observed spectrum $O_\lambda$ ({\it green}),
the model $M_\lambda$ ({\it red}) and the error ({\it blue}).
The {\it left-bottom} panel gives the residual spectrum $E_\lambda = O_\lambda - M_\lambda$.
Light-weighted stellar population fractions $x_{\rm j}$ are shown on the
{\it right-top} panel, on which the {\it left black} bar represents the power-law fraction
without a concrete age. Mass-weighted population fractions $\mu_{\rm j}$ are shown on the {\it right-bottom}
panel. We learn from Fig.~\ref{f5eps} that the stellar populations of the galaxy
consist of two main populations: a younger one $\leq$500\,Myr (whose 500\,Myr population is insignificant) 
and a mediate-to-old $>$1\,Gyr 
(we define young stellar populations as those with ages $\leq$300\,Myr, 
old populations with ages $>$1\,Gyr, and intermediate-age populations between them). 
Follow Cid Fernandes et al.\ (2005), we calculated the mean ages weighted by light,
\begin{equation}
\label{logtl}
\langle {\rm log} t_\star \rangle_L = \sum _{j=1}^{N_\star} x_j {\rm log} t_j
\end{equation}
and by stellar mass
\begin{equation}
\label{logtm}
\langle {\rm log} t_\star \rangle_M = \sum _{j=1}^{N_\star} \mu_j {\rm log} t_j
\end{equation}
The derived mean ages are $\langle {\rm log} t_\star \rangle_L = 8.2$
and $\langle {\rm log} t_\star \rangle_M = 9.7$.
These figures demonstrate again that the galaxy mass is dominated by old populations with ages
of $\sim 10^{10}$ years, and a latest starburst took place about $10^8$ years ago whose massive stars
contribute a large fraction of luminosity.
The mean velocity dispersion is $\sigma_\star = 223$ km s$^{-1}$, which is typical of giant ellipticals.

\subsection{Emission-Line Fitting}
\begin{figure*}[ht]
\begin{center}
\mbox{\subfigure{\includegraphics[width=80mm]{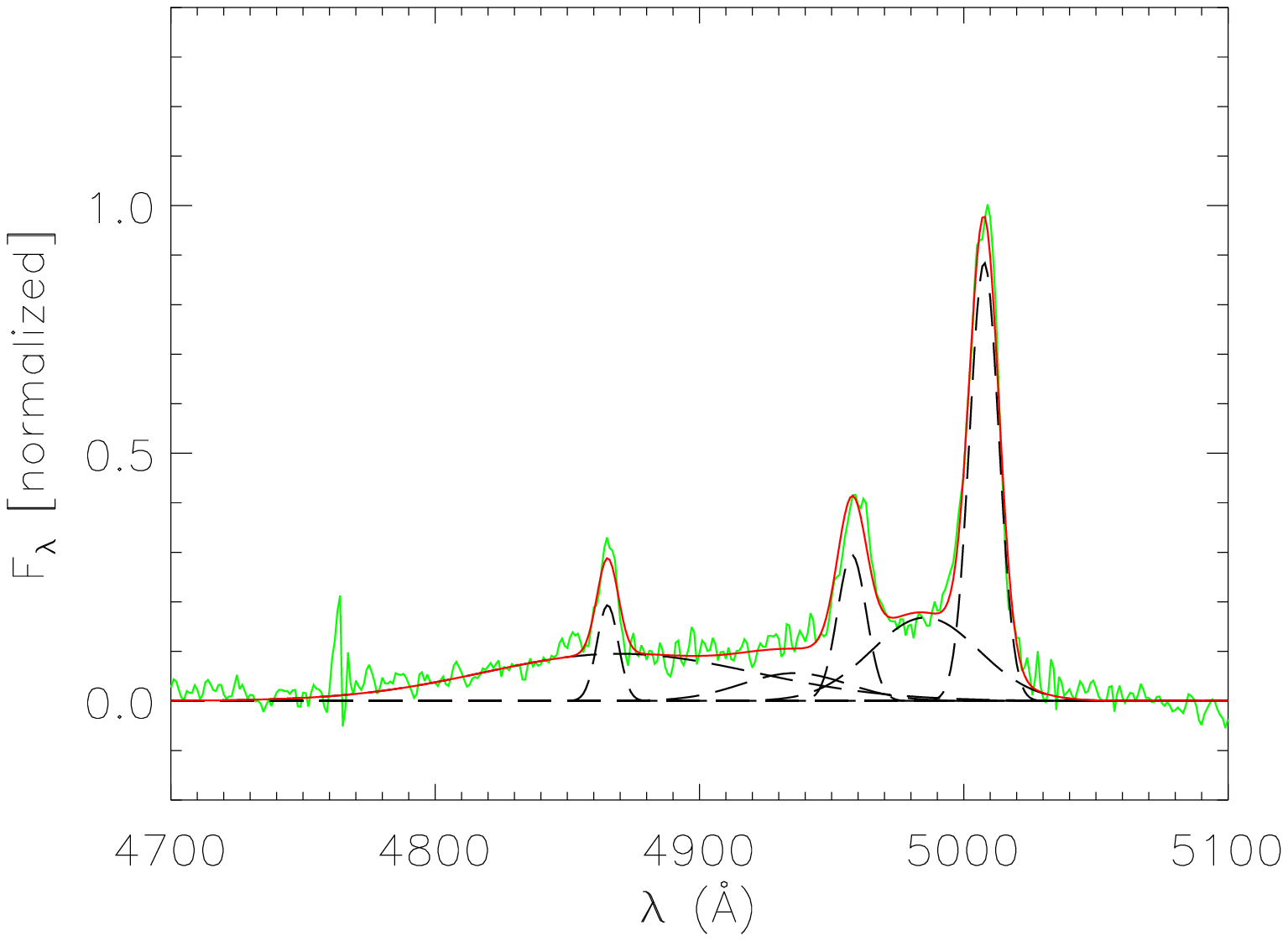}}\quad
\subfigure{\includegraphics[width=80mm]{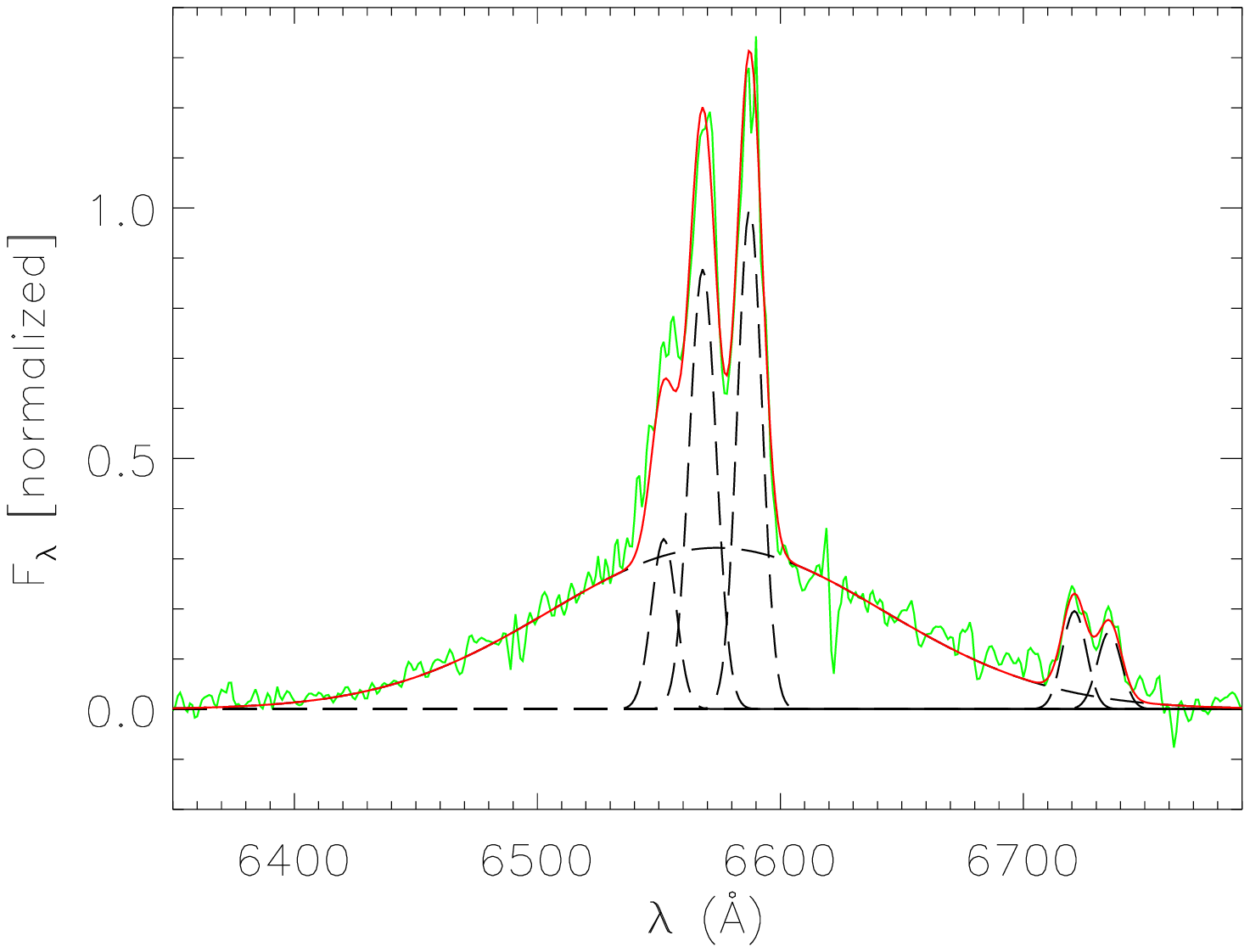}}\quad}
\caption{Fits to the emission lines. Flux intensities are inherited from the residual spectrum
in Fig.~\ref{f5eps}. The observed line profile ({\it green-solid}),
the model profile ({\it red-solid}),
each line and different components ({\it black-dashed}) are plotted in each panel.
{\it left}: H$\beta$+[O $_{\rm III}$] region; {\it right}: H$\alpha$+[N $_{\rm II}$]+[S $_{\rm II}$] region. }
\label{f6eps}
\end{center}
\end{figure*}

The residual spectrum $E_\lambda$ (the {\it purple} spectrum on the {\it left-bottom} panel) in Fig.~\ref{f5eps}
can be used to measure emission lines, since the stellar component and the AGN continuum have been removed. 
Fe $_{\rm II}$ emission regions are not included in emission-line fitting. 
Emission lines are modeled by the SPECFIT task in the IRAF-STSDAS package.
13 components are fitted simultaneously, including the single emission line [O $_{\rm II}$], 
[N $_{\rm II}$] and [S $_{\rm II}$] doublets, narrow and broad components 
of H$\beta$, H$\alpha$, [O $_{\rm III}$] $\lambda4959$ and  [O $_{\rm III}$] $\lambda5007$ 
(represented by H$\beta_{\rm{N}}$, H$\beta_{\rm{B}}$, and so on).
Each line as well as the narrow and broad components are fitted by a single Gaussian profile.
The flux ratios of [O $_{\rm III}$]$_N$, [O $_{\rm III}$]$_B$ and [N $_{\rm II}$] doublets
are fixed at their theoretical values. The relative positions of [O $_{\rm III}$],
[N $_{\rm II}$] and [S $_{\rm II}$] doublets are constrained by their laboratory values.
FWHMs are constrained to be same for each doublets like
[O $_{\rm III}$]$_N$, [O $_{\rm III}$]$_B$, [N $_{\rm II}$], [S $_{\rm II}$]
and also H$\beta$, H$\alpha$ narrow and broad components. The emission-line 
properties without aperture or extinction correction are displayed in Table 1. 
The uncertainties in columns (2) and (3) are given by the SPECFIT task.
Fitting results are shown in two wavelength ranges in Fig.~\ref{f6eps}. 
The narrow components of H$\alpha$ and H$\beta$ have FWHMs $<$1000 km s$^{\rm -1}$, 
and their broad components are $>$7000 km s$^{\rm -1}$. It can be classified as a quasar with
spectral type of Seyfert 1.5 galaxy, for both narrow and broad components of the two Balmer lines are significant.

Greene \& Ho (2005) found a tight correlation between Balmer emission-line luminosities 
and AGN optical continuum luminosity (little host galaxy contamination) 
at 5100\,\AA\ ($L_{5100} = \lambda L_\lambda$ at $\lambda = 5100$\,\AA).
We use this correlation to examine whether our spectral decomposition is successful.
We measure $L_{5100}$ from the power-law spectrum which is obtained from spectral synthesis.
The model AGN spectrum gives $L_{5100} = 10^{10.04}\,L_\odot$.
The observed H$\alpha_{\rm B}$ and H$\beta_{\rm B}$ luminosities are $L_{\rm H\alpha_{\rm B}} = 10^{8.73}\,L_\odot$
and $L_{\rm H\beta_{\rm B}} = 10^{8.07}\,L_\odot$.
Using Greene \& Ho's correlations, $L_{5100}$ is calculated to be $10^{10.06}\,L_\odot$ and 
$10^{9.97}\,L_\odot$, respectively. They are both consistent with the 5100\,\AA\ luminosity we modeled,
so the spectral decomposition of the AGN and host galaxy is reasonable.

\begin{table}[h]
\begin{center}
\caption{The emission-line properties of IRAS\,F13308+5946. Fluxes are measured within
the fiber aperture, without extinction correction, and in unit of
$4.549872\times10^{-16} {\rm (ergs~s^{-1}~cm^{-2})}$.}
\begin{tabular}{lccccc}
\tableline
Line Identification & Flux & FWHM$\rm{(km~s^{-1})}$\\
\tableline
$[\rm{O _{\rm II}}]\lambda$3727  &  3.96$\pm0.23$  &  725.85$\pm41.33$\\
H$\beta_{\rm{N}}$  &  1.96$\pm0.23$  &  587.42$\pm35.85$\\
H$\beta_{\rm{B}}$  &  12.17$\pm0.53$  &  7403.59$\pm97.26$\\
H$\alpha_{\rm{N}}$  &  12.01$\pm0.66$  &  587.42$\pm35.85$\\
H$\alpha_{\rm{B}}$  &  55.57$\pm1.06$   &  7403.59$\pm97.26$\\
$[\rm{O _{\rm III}}]\lambda4959_{\rm{N}}$    &  4.05$\pm0.16$  &  779.80$\pm30.76$\\
$[\rm{O _{\rm III}}]\lambda4959_{\rm{B}}$   &  2.83$\pm0.17$ &  2879.82$\pm184.07$\\
$[\rm{O _{\rm III}}]\lambda5007_{\rm{N}}$    &  12.26$\pm0.49$  &  779.80$\pm30.76$\\
$[\rm{O _{\rm III}}]\lambda5007_{\rm{B}}$   &  8.57$\pm0.51$ &  2879.82$\pm184.07$\\
$[\rm{N _{\rm II}}]\lambda6548$     &  4.29$\pm0.21$  &  543.60$\pm27.18$\\
$[\rm{N _{\rm II}}]\lambda6583$     &  12.69$\pm0.62$  &  543.60$\pm27.18$\\
$[\rm{S _{\rm II}}]\lambda$6716     &  2.41$\pm0.23$  &  516.52$\pm49.54$\\
$[\rm{S _{\rm II}}]\lambda$6731     &  1.91$\pm0.20$  &  516.52$\pm49.54$\\
\tableline
\end{tabular}
\end{center}
\end{table}

\subsection{Aperture Correction}
We have showed in Fig.~\ref{f2eps} that the observed spectrum is the flux from the 3\arcsec\ diameter fiber
of SDSS spectrograph. Precise calculations must include the flux come from the whole galaxy.
Table 2 lists the Galactic reddening corrected fiber magnitude ({\it fiberMag},
measured within the aperture of a fiber), Petrosian magnitude ({\it petroMag},
measured by the modified form of the Petrosian system) and {\it K}-correction factor for each band.
The aperture effect is significant,
for the difference between {\it fiberMag} and {\it petroMag} can be larger than 1 magnitude at some bands.
Since there's no lager aperture spectrum of the galaxy available by now, aperture correction is
necessary in this work.
A rough estimation of the aperture effect can be derived via
\begin{equation}
\label{aperture}
A = L_{\rm Petro}/L_{\rm fiber} = 10^{-0.4(m_{\rm Petro}-m_{\rm fiber})}
\end{equation}
(Hopkins et al.\ 2003). The correction factors for {\it u-, g-, r-, i-, z}-bands are 1.68, 2.2, 2.58, 2.60 and 2.57,
respectively. At {\it u}-band, the aperture effect is the least, but it is still a factor of $A = 1.68$.
From the spectral synthesis in \S3.3, it is known that the luminosity of the galaxy 
is dominated by young stellar populations and the AGN, 
which both can be better traced by {\it u}-band rather than other four bands, 
so we carry out aperture correction using {\it u}-band's factor.

\begin{table}[h]
\begin{center}
\caption{Magnitudes from SDSS fiber and Petrosian photometry.}
\begin{tabular}{lccccc}
\tableline
mag & u & g & r & i & z\\
\tableline
fiberMag & 18.46 & 17.86 & 17.37 & 17.06 & 16.84\\
petroMag & 17.90 & 17.00 & 16.34 & 16.02 & 15.82\\
k-correction & 0.028 & 0.166 & 0.109 & 0.0213 & 0.086\\
\tableline
\end{tabular}
\end{center}
\end{table}

Equation~(\ref{aperture}) is a simple aperture correction, because 
it doesn't distinguish different origins of the observed light. 
The 3\arcsec\ fiber is of two orders larger than typical AGN 
scale (for narrow-line region, $\sim$100 pc), so the 
3\arcsec\ fiber and the Petrosian radius may both embrace the 
whole AGN region. However, the stellar content varies in the two 
apertures (we see Petrosian radius a larger aperture). 
If we want to derive the flux of a starburst powered emission-line 
coming from the whole galaxy and adopt equation~(\ref{aperture}), 
we will amplify the AGN component and therefore underestimate the contribution from stars.

The spectral synthesis method provides us an opportunity to improve the aperture correction method above. 
Having decomposed the observed spectrum into AGN power-law and stellar component, 
we can convert their fluxes at a given band to apparent magnitudes. 
In combination with $m_{\rm fiber}$ and $m_{\rm Petro}$, 
we give a new method for purely stellar component aperture correction. The derived formula is:
\begin{equation}
\label{aperturefine}
A = {L_{\rm \star Petro} \over L_{\rm \star fiber}} = {{10^{-0.4(m_{\rm Petro}-m_{\rm AGN})}-1} \over {10^{-0.4(m_{\rm fiber}-m_{\rm AGN})}-1}}
\end{equation}
where $L_{\rm \star petro}$ and $L_{\rm \star fiber}$ are fluxes of stars at a given 
band within Petrosian and fiber aperture, respectively; $m_{\rm AGN}$ is the apparent magnitude of AGN 
at a given band converted from the flux density of the power-law spectrum. 

\subsection{Extinction Curve}
Large extinction happens in ULIRGs/LIRGs and re-emits UV-to-optical emission to FIR. 
The proper dust extinction law we use here is the one given by Calzetti et al.\ (1994; 1997; 2000) 
and Leitherer et al.\ (2002). 
Calzetti curve is only applicable to wavelengths longer than 1200\AA. 
Leitherer et al.\ (2002) extended the curve to 970\AA. At even shorter wavelengths, 
extinction curve has larger systematic uncertainties, but we extrapolate it a little to 912\AA\ 
to cover the Lyman series limit. 
Their curves are identical at 1500\AA, and the differences at shortward wavelengths are minor. 
We adopt Leitherer et al.'s curve for 912\AA $\leq \lambda \leq$ 1800\AA, 
and adopt Calzetti curve for 1800\AA $< \lambda \leq$ 9000\AA.

The intrinsic emission $F_i(\lambda)$ can be recovered through the extinction curve $k^{'}(\lambda)$ via
\begin{equation}
\label{intrinflux}
F_{\rm i}(\lambda) = F_{\rm o}(\lambda)10^{0.4[E_s(B\!-\!V) k^{'}(\lambda) - E_s(B\!-\!V) k^{'}_V]}
\end{equation}
where $F_o(\lambda)$ is the observed spectrum,
\esbv\ is the color excess for the stellar and AGN continuum spectrum,
$10^{0.4[E_s(B\!-\!V) k^{'}(\lambda) - E_s(B\!-\!V) k^{'}_V]}$ corresponds to
the global extinction term $r_\lambda$ in equation~(\ref{Mlambda}).
\esbv\ is calculated from $A_{\rm v}$ obtained from the spectral
synthesis ($A_{\rm V} = 0.62$) and it is $\esbv = 0.15$.
The color excess for nebular gas emission-lines is denoted by \egbv\ and is directly estimated through
Balmer decrement, the line ratio ${\rm H\alpha_N/H\beta_N} = 6.13$, whose intrinsic flux
ratio is $({\rm H\alpha_N/H\beta_N})_0 = 2.87$ assuming temperature $T = 10000$ K and case B
recombination (Osterbrock 1989). \egbv\ is derived to be $\egbv = 0.71$.

When applying the Calzetti curve to B-band,
the extinction correction factor is $\sim$4, corresponding to 1.5
magnitude brighter, such the corrected absolute magnitude 
is $M_B = -22.6$, which satisfies the quasar criterion $M_B < -22.2$.

\section{SFR and Black-Hole Mass}
We can compare the observed IR luminosity with the current SFR. 
Because the H$\alpha$ line contains emission from the AGN, we use [O $_{\rm II}$] $\lambda3727$ 
to estimated the SFR. We take color excess $\egbv = 0.71$ for gas reddening and Calzetti curve 
for spectral line extinction correction. The current SFR is calculated by
\begin{eqnarray}
\label{SFROII}
{\rm SFR([O~_{II}])}(M_\odot\,{\rm yr^{-1}}) &=& (6.58 \pm 1.65) \nonumber \\
                                        & & \times 10^{-42}\,L_{[{\rm O~_{II}}]} ({\rm {ergs~s^{-1}}})
\end{eqnarray}
(Kewley et al.\ 2004), where $L_{[{\rm O~_{II}}]}$ is the extinction corrected 
(a factor of 45.6) luminosity of [O $_{\rm II}$] $\lambda3727$. 
The derived SFR is 43.8$\pm$11.0\,$M_\odot$\,yr$^{-1}$. We don't take aperture correction here, 
because we have learnt that the starburst takes place mainly at the center region of the galaxy (see \S2.2), 
where the fiber aperture covers. Considering the star-forming region can be also found 
outside the aperture, we take such SFR as a lower-limit and the {\it u}-band aperture corrected SFR 
as an upper-limit, which is 73.5$\pm$18.4\,$M_\odot$\,yr$^{-1}$. 
For starburst galaxies, SFR can be also estimated through IR luminosity (Kennicutt 1998, and references therein). 
As for IRAS\,F13308+5946, since it contains both starburst and AGN components, 
the total IR luminosity $10^{11.56}\,L_\odot$ only suggests an upper limit of SFR, which is 62.9\,$M_\odot$\,yr$^{-1}$. 
The spectral synthesis has given the starburst fraction of the galaxy, so the SFR can be 
estimated only by the IR luminosity from the starburst. Such calculation will be given in \S5. 

The BH virial mass is calculated by means of H$\alpha_{\rm B}$ via
\begin{eqnarray}
\label{MBH}
M_{\rm BH} &=& (2.0_{-0.3}^{+0.4}) \times 10^6 \biggl( {L_{\rm H\alpha_{\rm B}}\over 10^{42} {\rm ergs~s^{-1}}} \biggr)^{0.55\pm0.02} \nonumber \\
           & & \times \biggl( {{\rm FWHM_{H\alpha}}\over10^3 {\rm km~s^{-1}}}\biggr)^{2.06\pm0.06}\,M_\odot
\end{eqnarray}
(Greene \& Ho 2005). Where $L_{\rm H\alpha_{\rm B}}$ is the observed H$\alpha_{\rm B}$ luminosity.
We don't carry out aperture correction on $L_{\rm H\alpha_{\rm B}}$,
because the 3\arcsec\ fiber aperture covers $\sim$10 kpc, including the broad-line region definitely. 
The derived BH mass is $M_{\rm BH} = 1.8 \times 10^8\,M_\odot$. The Eddington ratio
$L_{\rm bol}/L_{\rm Edd}$ is calculated adopting the bolometric luminosity
$L_{\rm bol} \approx 9 \lambda L_\lambda$(5100\AA) (Kaspi et al.\ 2000),
where $L_\lambda$(5100\AA) is the AGN optical continuum luminosity at 5100\,\AA.
The inferred Eddington ratio is $L_{\rm bol}/L_{\rm Edd} = 0.12$,
which is typical of PG QSOs (median value 0.24; Hao et al.\ 2005, hereafter Hao05) 
and far less than the one of IR QSOs (1.73) and NLS1s (1.27). 
These results indicate the SMBH is at the late stage of growth.
If the SMBH grew from a single seed, the duration of the growth can be estimated by the ${\it e}$-folding timescale,
\begin{equation}
\label{tacc}
t_{\rm acc} \equiv  {M_{\rm BH}\over \dot{M}} = 4 \times 10^7 \Bigl( {\epsilon \over 0.1} \Bigr) \eta^{-1}\,{\rm yr}
\end{equation}
(Haiman \& Loeb 2001), where $\epsilon \equiv L_{\rm bol} / {\dot{M} c^2}$ is the mass-to-radiation
conversion efficiency and $\eta \equiv L_{\rm bol}/L_{\rm Edd}$ is the Eddington ratio. For efficient
energy production, $\epsilon \sim 0.1$ and the ${\it e}$-folding timescale
is $t_{\rm acc} = 3.5 \times 10^8$\,yr. It is close to the duration of the starburst phase. 
This means that before the starburst activity took place, the BH mass was less by a
factor of $\sim$2.7, which was still massive enough to support a broad-line region. 
Therefore, the galaxy was probable to harbor a type I AGN during the starburst phase.

\section{FIR Luminosity in the Past}
We have little information to determine the origin of the BH,
that is, whether it grew from a single seed or from
BHs merger when galaxies merged. Therefore, the AGN emission
in the past is hard to estimate. Comparing with the uncertainty
of AGN, we have better defined stellar evolution models (Kurucz
1992; Lejeune et al.\ 1997, 1998; BC03). Combining with the star
formation history, we are able to trace back to the past and
calculate the luminosity of the stellar component at a given
epoch. Then, we could estimate the IR luminosity by calculating
the UV-to-optical flux absorbed by dust. Because there is no UV
observation data available for this galaxy and the shortward end
of the rest-frame spectrum is only limited to 3258 {\rm \AA}, 
we use model spectrum of UV-band for such calculation.

\begin{figure}[htb]
\includegraphics[width = 84mm]{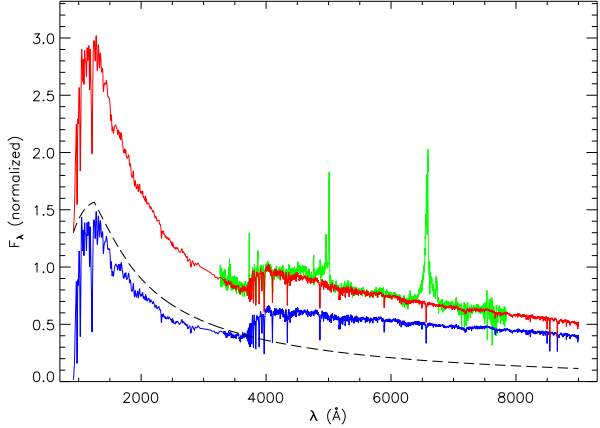}
\caption{The reconstructed model spectrum ({\it red-solid}) of the present state. The observed
spectrum ({\it green-solid}) is superimposed on it. The {\it blue-solid} line represents the
host galaxy starlight, and the {\it black-dashed} line represents the double power-law spectrum.}
\label{f7eps}
\end{figure}

To test the feasibility of our method, we firstly calculate the present FIR luminosity 
by reconstructing the UV-to-optical spectrum. 
Because BC03 SSP templates are in units of solar luminosity per angstrom per solar mass,
we can accurately convert the present stellar mass ${\rm Mcor\_tot}$ to a spectrum
by means of the mass weighted fraction $\mu_{\rm j}$. We reconstruct the model spectrum
$F_{\rm o}(\lambda,t_0)$ covering the whole UV-to-optical wavelengths in our calculation
(from 912{\rm \AA} to 9000{\rm \AA}) through
\begin{equation}
\label{rebuiltp}
F_{\rm o}(\lambda,t_0) = [{\rm Mcor\_tot} \sum _{\rm j=1}^N {\mu_{\rm j} B_{\lambda,\rm j,t_0}+F_p(\lambda)] 10^{-0.4(A_\lambda-A_V)}}
\end{equation}
where $t_0$ is the present time, ${\rm Mcor\_tot}$ is the present stellar mass
obtained from the spectral synthesis and is derived to be $5.5 \times 10^{10}\,M_\odot$
after aperture correction through equation~(\ref{aperturefine}) (doesn't need extinction correction), 
$B_{\lambda,\rm j,\rm t_0}$ are BC03 SSP templates without normalization,
$F_p(\lambda)$ is a double power-law spectrum of AGN. The spectral indices of power-law we use here
are given by Hatziminaoglou et al.\ (2008) that $\alpha = -1$ for $\lambda < 1250$\AA\ 
and $\alpha = -2$ for $\lambda \geq 1250$\AA. 
Fig.~\ref{f7eps} shows the reconstructed model spectrum ({\it red-solid}) superimposed by the
observed one ({\it green-solid}). This model spectrum is used as $F_o(\lambda)$ in equation~(\ref{intrinflux}).

The light in UV-to-optical (912--9000\AA) absorbed by dust and re-emitted to IR is approximately calculated by
\begin{equation}
\label{Lirt}
L_{{\rm IR},t} = \int _{912}^{9000} \bigl[F_{\rm i}(\lambda, t) - F_{\rm o}(\lambda, t)\bigr]{\rm d}\lambda
\end{equation}
where $F_{\rm i}(\lambda, t)$ and $F_{\rm o}(\lambda, t)$ have similar meaning
as $F_{\rm i}(\lambda)$ and $F_{\rm o}(\lambda)$ in equation~(\ref{intrinflux}), but they 
represent the spectra at look-back time {\it t}. After {\it u}-band
aperture correction, the derived IR luminosity is $L_{{\rm IR},0}=10^{11.43}\,L_{\odot}$.
It approximates to the observed $L_{\rm IR}=10^{11.56}\,L_{\odot}$.
It demonstrates that our method is feasible in recovering the present FIR luminosity,
so we can apply the method to the past.
Meanwhile, it is convenient to calculate the contribution fractions from starburst and AGN.
We only need to calculate IR luminosities from the two components separately. 
The starburst and AGN contribute $L_{\rm star,IR}=10^{11.28}\,L_{\odot}$ and
$L_{\rm AGN,IR}=10^{10.90}\,L_{\odot}$, respectively.
The starburst contributes $\sim70\%$ of the FIR luminosity, nearly
two thirds of the total FIR budget. Anyway, two components are both important.

The current SFR can be estimated through the relation between $L_{\rm FIR}(8\text{--}1000\mu{\rm m})$ and SFR (Kennicutt 1998). 
Since the host galaxy contributes $\sim70\%$ of the FIR luminosity, we obtain ${\rm SFR} = 43.3\,M_\odot$\,yr$^{-1}$. 
It is consistent with the one from [O $_{\rm II}$] luminosity inside the aperture. 
Therefore, the derived stellar populations and AGN spectra, the dust extinction and the observed emission
line fluxes are concordant with each other.

The estimation of the past IR luminosity is carried out in a
similar way. We just need to supplement the stellar mass loss of
each population at a given look-back time {\it t}. Because the AGN
emission in the past is unknown, we just estimate the stellar
component contribution. The stellar spectrum (i.e. the host
galaxy) in the past without extinction is produced through
\begin{equation}
\label{youngp}
F_{\rm i}({\lambda,t}) = {\rm Mcor\_tot} \sum _{{\rm j}=1}^{\rm N} {{\mu_{\rm j} \over f_{\star,{\rm j},t_0}} f_{\star,{\rm j},t} B_{\lambda,{\rm j},t}}
\end{equation}
where $f_{\star,{\rm j},t_0}$ is the present ($t_0$) fraction of the remaining stellar mass 
to the initial mass of population {\it j}, $f_{\star,{\rm j},t}$ is such fraction 
at a given time {\it t}, $f_{\star,{\rm j},t_0}$ and $f_{\star,{\rm j},t}$ 
are in the range 0 $< f_{\star,{\rm j},t_0}, f_{\star,{\rm j},t} \leq 1$. 
From Fig.~\ref{f5eps} (the {\it bottom-right} panel) 
Since the recent starburst happened in the past 500\,Myr, we estimate 
the IR luminosity of the past 25, 100, 290 and 500\,Myr separately.

\begin{figure}[htb]
\includegraphics[width = 84mm]{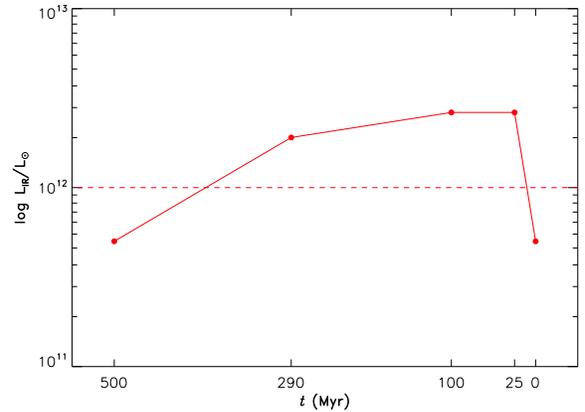}
\caption{The probable IR luminosities during the starburst phase. The ULIRG phase may lasted for $\sim$300\,Myr.}
\label{f8eps}
\end{figure}

Take 25\,Myr ago as an example. When the 25 old population was just newly born, 
it was assigned an age of 1\,Myr and the parameter $f_{\star,{\rm j},t}$ is equal to 1. 
The 100\,Myr old population was 75 old then with $f_{\star,{\rm j},t} = 0.7226$. 
Other populations are also configured in the same way. 
Here, the reconstructed spectrum through equation~(\ref{youngp}) is used as the intrinsic 
spectrum $F_{\rm i}(\lambda)$ in equation~(\ref{intrinflux}). 
The extinction term $A_{\rm \lambda}$ is assigned to be as the present. 
The attenuated spectrum through $F_{\rm i}(\lambda, t) 10^{-0.4 A_{\rm \lambda}}$ 
is used as $F_{\rm o}(\lambda)$ in equation~(\ref{intrinflux}). 
Because dust extinction increases from optical to UV, 
we adopt the aperture correction at {\it u}-band derived from equation~(\ref{aperturefine}). 
Calculated through equations~(\ref{intrinflux}) and~(\ref{Lirt}), 
the IR luminosity transferred from 912--9000\AA\ to IR at 25\,Myr ago 
was $L_{{\rm IR, 25M}}=10^{12.18}\,L_{\odot}$, which is ULIRG luminosity. 
If we reckon in the AGN's contribution, the IR luminosity should be higher. 
When applying to 100, 290 and 500\,Myr ago, the estimated IR luminosities from the host galaxy 
were $L_{{\rm IR, 100M}}=10^{12.19}\,L_{\odot}$, $L_{{\rm IR, 290M}}=10^{12.11}\,L_{\odot}$ 
and $L_{{\rm IR, 500M}}=10^{11.45}\,L_{\odot}$. 
These IR luminosities are shown as a history in Fig.~\ref{f8eps}. 
As has been demonstrated in section \S4, the galaxy was probable to harbor a type I AGN during the 
starburst. Therefore, a type I ULIRG might appeared during the starburst phase and lasted for $\sim$300\,Myr. 

\section{Discussion}
\subsection{Fe $_{\rm II}$ pseudo-continuum}
Many IR QSOs are extremely strong Fe $_{\rm II}$ emitters (Zheng et al.\ 2002; L\'ipari et al.\ 2003). 
In IRAS\,F13308+5946 spectrum, we also find Fe $_{\rm II}$ 
pseudo-continuum emission in wavelength region 4270--4700\AA\ and 
5100--5600\AA. We have masked these two wavelength regions before spectral synthesis. 
Actually, we have attempted to subtract Fe $_{\rm II}$ lines from $O_\lambda$ before 
fitting. However, as the case of IRAS\,Z11598$-$0112, IRAS\,F02065+4705, etc.\ (Zheng 
et al.\ 2002), Fe $_{\rm II}$ multiplets 37, 38 (4500--4680\AA) are relatively 
stronger than the Fe $_{\rm II}$ multiplets 48, 49 (5100--5400\AA) 
compared with Boroson \& Green (1992, hereafter BG92) Fe $_{\rm II}$ template, 
and also the V\'eron-Cetty et al.\ (2004) template. 
Thus, the Fe $_{\rm II}$ multiplets 37, 38 are left when 
the multiplets 48, 49 are removed. We have also attempted to 
remove Fe $_{\rm II}$ lines from the residual spectrum, but the 
result was not improved. Furthermore, including Fe $_{\rm II}$ template as a 
component to fit $O_\lambda$ does not bring any improvement at all. 
Therefore, we masked the Fe $_{\rm II}$ pseudo-continuum regions before spectral synthesis, 
as have mentioned in \S3.4.

The Fe $_{\rm II}$ $\lambda$4570 (4434--4684\AA) flux is measured during the procedures 
above and the line ratio Fe $_{\rm II}~\lambda$4570/H$\beta$ is 0.81. 
This ratio is typical of optical selected QSOs (BG92). 
We find that the Fe $_{\rm II}$ continuum extension to [O $_{\rm II}$] $\lambda3727$, 
H$\beta$, etc.\ lines is hard to detect, so we measure these lines fluxes directly.

\subsection{Extinction Curve}
In this paper we don't adopt Cardelli et al.'s (1989) extinction curve, 
because it is for cases which are similar to Milky Way and not applicable to starburst galaxies. 
We use the Calzetti curve and Leitherer et al.'s (2002) extension form, 
because they are derived from star-forming regions and starburst galaxies. 

Dust in different components and stellar populations vary 
in temperature, ingredient, amount, geometry, etc., so different
attenuation for each component should be adopted if we could
distinguish them. However, as the spectra from the host galaxy and central AGN are coupled, 
and also there is only one extinction term in \starlight, the extinction 
obtained from spectral synthesis is a combination value from both regions. 
Also, there may exist some optically 
thick star formation regions, which may not be detected even by 
optical or near-infrared, but they still contribute to IR flux. In this case, the 
actual extinction could be underestimated.

From Balmer decrement, the line ratio H$\alpha/$H$\beta$, the derived
color excess of stellar continuum  is $\esbv = 0.3115$,
which is twice as the one obtained from spectral fitting. This
indicates differential attenuation between the dusty starburst
region and older populations.

\subsection{IRAS\,F13308+5946: A Possible Evolutionary Link between (type I) ULIRGs and QSOs}
The spectral synthesis gives a 500\,Myr starburst history, 
which implies a triggering mechanism took place during such epoch. 
Galaxy interactions and mergers have timescales of a few $\times 10^8$\,yr 
(Binney \& Tremaine 1987). 
The similar timescales imply that the starburst was possibly triggered by galaxies merger.

\begin{figure}[h]
\includegraphics[width = 84mm]{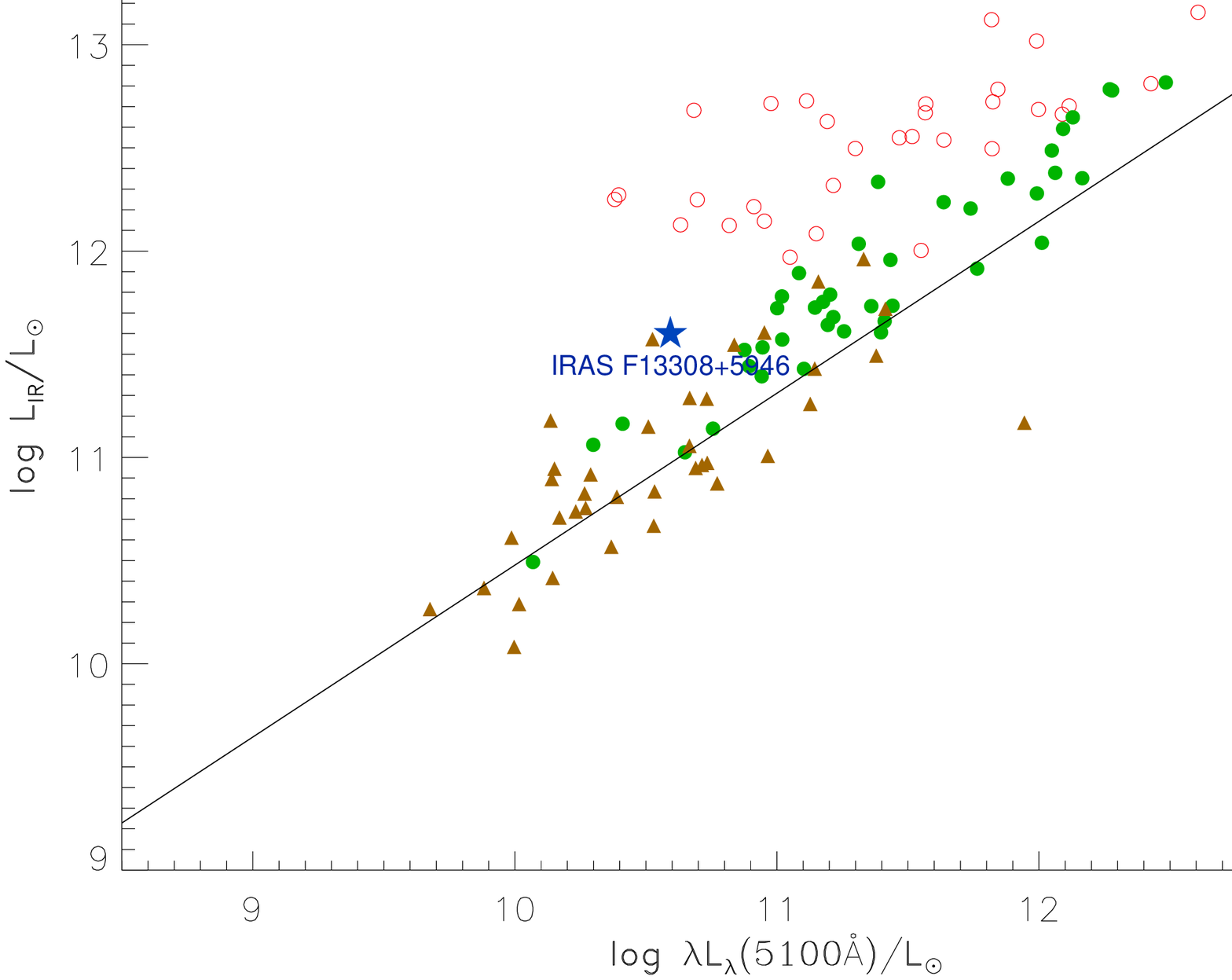}
\caption{IR luminosity vs. the bolometric luminosity measurement, $\lambda L_\lambda$5100\AA.
The {\it red} open circles represent IR QSOs; the {\it green} filled circles
represent PG QSOs; the {\it brown} filled triangles represent NLS1s;
the solid line represents the linear regression for all PG QSOs and NLS1s.
These data points are reproduced with the data from Hao05.
IRAS\,F13308+5946 is represented by the {\it blue} star.}
\label{f9eps}
\end{figure}

\begin{figure}[h]
\includegraphics[width = 84mm]{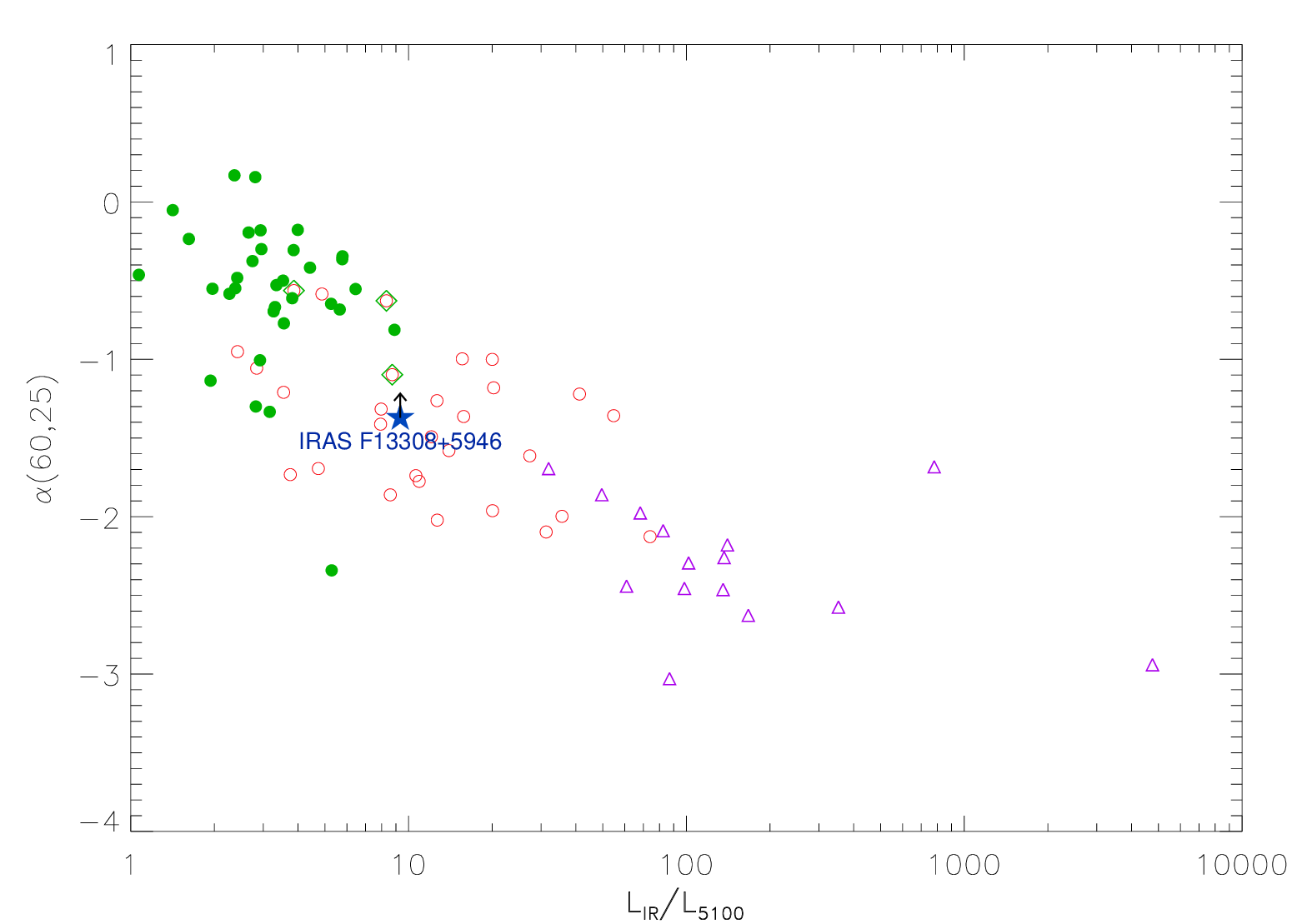}
\caption{IR spectral index $\alpha(60, 25)$ vs. the IR excess, $L_{\rm IR}/L_{\rm 5100}$.
The {\it purple} open triangles represent 15 ULIRGs, which are selected from Kim et al.\ (1998b)
and have SDSS spectra; other data points are also from Hao05 and have the same meaning as Fig.~\ref{f9eps};
the three {\it red} open circles enclosed by {\it green} diamonds represent IR QSOs which are also PG QSOs.
IRAS\,F13308+5946 is represented by the {\it blue} star and marked with a lower limit.}
\label{f10eps}
\end{figure}

In \S4 and \S5, we demonstrated that IRAS\,F13308+5946 has possibly experienced 
an IR QSO phase since $\sim$300\,Myr ago. Follow Hao05, 
we plot IRAS\,F13308+5946 on the diagram which shows the relation 
between IR luminosity and the bolometric luminosity measurement 
(represented by $\lambda L_\lambda(5100\AA)$; Kaspi et al.\ 2000) in Fig.~\ref{f9eps}. 
IRAS\,F13308+5946 locates at a transitional position between IR QSOs and PG QSOs. 
The tight correlation followed by PG QSOs and NLS1s suggests their IR luminosities 
are associated with the optical through central AGNs, 
while IR QSOs have IR excess powered by other mechanisms rather than AGN. 
As for IRAS F13308+5946, we attribute the IR excess to the ongoing star formation. 
In Fig.~\ref{f10eps}, which shows the IR spectral index $\alpha(60, 25)$ 
(defined as $\alpha(\lambda_1, \lambda_2)=-{{\rm log}[F(\lambda_2)/F(\lambda_1)] \over {\rm log}(\lambda_2/\lambda_1)}$) 
versus the IR excess $L_{\rm IR}/L_{\rm 5000}$, it also locates between IR QSOs and PG QSOs. 
The IR excess increase as the dust temperature decrease, suggesting that star formation 
activities become more and more important in powering IR luminosity of IR QSOs and ULIRGs, 
because the temperature of dust heated by stars is lower than that by AGN. 

On the other hand, Hao05 gives statistical median $M_{\rm BH}$ values of IR QSOs and PG QSOs, 
which are 4.9$\times10^7\,M_\odot$ and 2.1$\times10^8\,M_\odot$. 
Thus, IRAS\,F13308+5946 has both BH mass and Eddington ratio similar to those of PG QSOs. 
Furthermore, as a type 1.5 galaxy, the featureless continuum and emission-lines 
from the AGN may be partially attenuated by the dusty torus, and the AGN itself may be powering a classical QSO, 
for the galaxy is an {\it i}-band quasar. 

Therefore, IRAS\,F13308+5946 may evolve into an optical QSO when the
starburst ceases and the nuclear dust is dissipated by the radiative pressure of AGN. 
In Fig.~\ref{f9eps} and \ref{f10eps}, the positional arrangement of ULIRGs, IR QSOs and classical QSOs implies an
evolutionary sequence: ULIRGs $\rightarrow$ type I ULIRGs / IR QSOs $\rightarrow$ (IRAS\,F13308+5946) $\rightarrow$ PG QSOs. 
As the {\it H}-band luminosity is $\sim$4$L$*, it support the evolutionary scenario proposed 
by Colina et al.\ (2001) that ULIRGs generated by two or more massive ($\geq L_*$) galaxies would
evolve into QSOs (also supported by Sanders et al.\ 1988 and Lutz et al.\ 1999).

Finally, through {\it r}-band surface photometry, we find the outskirt surface brightness
profile follows the de Vaucouleurs law and the central region follows a point spread function (PSF).
It is consistent with other studies which show that some (maybe even higher fraction)
ULIRGs may undergo QSOs phase in their evolutionary history before they settle down
as ellipticals (Zheng et al.\ 1999; Arribas et al.\ 2004; Dasyra et al 2006b).

\section{Summary}
We carry out a study based on stellar population synthesis result of a type I 
LIRG, IRAS\,F13308+5946. We find that:

1. The cross-identification from {\it IRAS}, 2MASS, SDSS and FIRST, combining with the
radio-FIR correlation, confirms that the LIRG IRAS\,F13308+5946 is the optical quasar
from SDSS observation.

2. With sub-solar metallicity Z=0.008 and power-law index $\alpha=-2.0$,
stellar population synthesis shows that the host galaxy has a recent starburst
history since 500\,Myr ago.

3. We estimate the past IR luminosity during the starburst epoch.
We find it has probably experienced a type I ULIRG phase from $\sim300$\,Myr ago.
When the star formation activity weakened recently, the IR luminosity decrease to the present
level ($10^{11.56}\,L_{\odot}$) as a LIRG.
Nuclear starburst and AGN activity both contribute to the IR luminosity budget, with $\sim$70\% from starburst.

4. The SMBH mass $M_{\rm BH}=1.8\times10^8 M_\odot$ and the Eddington ratio 
$L_{\rm bol}/L_{\rm Edd} = 0.12$ are both consistent with PG QSOs, suggesting a potential classical QSO.

5. It locates at the transitional position between IR QSOs and PG QSOs on the plots
$L_{\rm IR}$ versus $\lambda L_\lambda {\rm 5100}$ and $\alpha(60, 25)$
versus $L_{\rm IR}/L_{\rm 5100}$. Combining the results above, we conclude that IRAS\,F13308+5946 is probable 
an evolutionary transition object from a type I ULIRG to a PG QSO.

\acknowledgments
The authors are grateful to Dr.\ Cai-Na Hao for her data through private communication.
We thank R.\ Cid Fernandes for his warm-hearted explanation of \starlight\ and very valuable
advices to the paper. We thank Dr.\ Stijn Wuyts for valuable discussions.
This project is supported by NSFC grant 10833006, 10773014 and the 973 Program grant 2007CB815406.

This work is based on the data from FIRST, IRAS, 2MASS and SDSS observations.
Funding for the SDSS and SDSS-II has been provided by the Alfred P. Sloan Foundation,
the Participating Institutions, the National Science Foundation, the U.S.\ Department of Energy,
the National Aeronautics and Space Administration, the Japanese Monbukagakusho,
the Max Planck Society, and the Higher Education Funding Council for England.
The SDSS Web Site is http://www.sdss.org/.
The 2MASS project is a collaboration between The University of Massachusetts and the
Infrared Processing and Analysis Center (JPL/Caltech). Funding is provided primarily by
NASA and the NSF. The University of Massachusetts constructed and maintained the observatory
facilities, and operated the survey. All data processing and data product generation is being
carried out by IPAC. Survey operations began in Spring 1997 and concluded in Spring 2001.

\clearpage


\begin{thebibliography}{}
  \bibitem[1993]{Alongi1993} Alongi M., Bertelli G., Bressan A., Chiosi C., Fagotto F., Greggio L., NasiE., 1993, A\&AS, 97, 851
  \bibitem[2003]{Arribas2003} Arribas, S., \& Colina, L. 2003, ApJ, 591, 791
  \bibitem[2004]{Arribas2004} Arribas, S., Bushouse, H., Lucas, R. A., Colina, L., \& Borne, K. D. 2004, AJ, 127, 2522
  \bibitem[1987]{Binney1987} Binney, J., \& Tremaine, S. 1987, Galactic Dynamics (Princeton: Princeton Univ. Press)
  \bibitem[2003]{Blanton2003} Blanton, M. R., et al.\ 2003, AJ, 125, 2348
  \bibitem[2002]{Boller2002} Boller, T., Gallo, L. C., Lutz, D., \& Sturm, E. 2002, MNRAS, 336, 1143
  \bibitem[1992]{Boroson1992} Boroson, T. A., \& Green, R. F. 1992, ApJS, 80, 109 (BG92)
  \bibitem[1993]{Bressan1993} Bressan A., Fagotto F., Bertelli G., Chiosi C., 1993, A\&AS, 100, 647
  \bibitem[2003]{Bruzual2003} Bruzual G., \& Charlot S., 2003, MNRAS, 344, 1000 (BC03)
  \bibitem[2002]{Bushouse2002} Bushouse H. A., Borne K. D., Colina L. et al.\ 2002, ApJS, 138, 1
  \bibitem[1997]{Calzetti1997} Calzetti, D. 1997, in AIP Conf. Proc. 408, The Ultraviolet Universe at Low and High Redshift: Probing the Progress of Galaxy Evolution, eds. W. H. Waller, M. N. Fanelli, J. E. Hollis, \& A. C. Danks (Woodbury: AIP), 403
  \bibitem[2000]{Calzetti2000} Calzetti, D., Armus, L., Bohlin, R.C. et al., 2000. ApJ 533, 682
  \bibitem[1994]{Calzetti1994} Calzetti, D., Kinney, A.L., Storchi-Bergmann, T., 1994. ApJ 429,582
  \bibitem[2000]{Canalizo2000} Canalizo, G., \& Stockton, A. 2000, AJ, 120, 1750
  \bibitem[2001]{Canalizo2001} Canalizo, G., \& Stockton, A. 2001, ApJ, 555, 719
  \bibitem[2006]{Cao2006} Cao, C., Wu, H., Wang, J. L., Hao, C. N., Deng, Z. G., Xia, X. Y., \& Zou, Z. L. 2006, Chinese J. Astron. Astrophys., 6, 197
  \bibitem[1989]{Cardelli1989} Cardelli, J. A., Clayton, G. C., \& Mathis, J. S. 1989, ApJ, 345, 245
  \bibitem[2003]{Chabrier2003} Chabrier G., 2003, PASP, 115, 763
  \bibitem[2005]{Cid Fernandes2005} Cid Fernandes R., Mateus A., Sodr\'e L., Stasi\'nska G., Gomes J. M., 2005, MNRAS, 358, 363
  \bibitem[1996]{Clements1996} Clements, D. L., Sutherland, W. J., McMahon, R. G., \& Saunders, W. 1996, MNRAS, 279, 477
  \bibitem[2001]{Colina2001} Colina, L., et al.\ 2001, ApJ, 563, 546
  \bibitem[2006]{Dasyra2006a} Dasyra, K. M., et al.\ 2006a, ApJ, 651, 835
  \bibitem[2006]{Dasyra2006b} Dasyra, K. M., Tacconi, L. J., Davies, R. I., Genzel, R., Lutz, D., Naab, T., Sanders, D. B., Veilleux, S., \& Baker, A. J. 2006b, New Astronomy Review, 50, 720
  \bibitem[1998]{Downes1998} Downes, D., \& Solomon, P. M. 1998, ApJ, 507, 615
  \bibitem[1994]{Fagotto1994a} Fagotto F., Bressan A., Bertelli G., Chiosi C., 1994a, A\&AS, 104, 365
  \bibitem[1994]{Fagotto1994b} Fagotto F., Bressan A., Bertelli G., Chiosi C., 1994b, A\&AS, 105, 29
  \bibitem[2003]{Franceschini2003} Franceschini, A., et al.\ 2003, MNRAS, 343, 1181
  \bibitem[1996]{Francis1996} Francis, P. J. 1996, Publ. Astron. Soc. Australia, 13, 212
  \bibitem[2001]{Genzel2001} Genzel, R., Tacconi, L. J., Rigopoulou, D., Lutz, D., \& Tecza, M. 2001, ApJ, 563, 527
  \bibitem[1996]{Girardi1996} Girardi L., Bressan A., Chiosi C., Bertelli G., Nasi E., 1996, A\&AS, 117, 113
  \bibitem[2005]{Greene2005} Greene, J. E., \& Ho, L. C., 2005, ApJ, 630, 122
  \bibitem[1997]{Gu1997} Gu, Q. S., Huang, J. H., Su, H. J., \& Shang, Z. H. 1997, A\&A, 319, 92
  \bibitem[2001]{Haiman2001} Haiman, Z. \& Loeb, A. 2001, ApJ, 552, 459
  \bibitem[2005]{Hao2005} Hao C. N., Xia X. Y., Mao S., Wu H., Deng Z. G., 2005, ApJ, 625, 78
  \bibitem[2008]{Hatziminaoglou2008} Hatziminaoglou, E. et al.\ 2008, MNRAS, 386, 1252
  \bibitem[2003]{Hopkins2003} Hopkins, A. M., et al.\ 2003, ApJ, 599, 971
  \bibitem[2009]{Hou2009} Hou, L. G., Wu, Xue-Bing, \& Han, J. L., 2009, ApJ, 704, 789
  \bibitem[2005]{Jester2005} Jester, S., et al.\ 2005, AJ, 130, 873
  \bibitem[2000]{Kaspi2000} Kaspi, S., Smith, P. S., Netzer, H., Maoz, D., Jannuzi, B. T., \& Giveon, U. 2000, ApJ, 533, 631
  \bibitem[2006]{Kawakatu2006} Kawakatu, N., Anabuki, N., Nagao, T., Umemura, M., \& Nakagawa, T. 2006, ApJ, 637, 104
  \bibitem[1998]{Kennicutt1998} Kennicutt, R. C., Jr. 1998, ARA\&A, 36, 189
  \bibitem[2004]{Kewley2004} Kewley, L. J., Geller, M. J., \& Jansen, R. A. 2004, AJ, 127, 2002
  \bibitem[2001]{Kewley2001} Kewley, L. J., Heisler, C.A., Dopita, M.A., Lumsden, S., 2001, ApJS, 132, 37
  \bibitem[1998]{Kim1998b} Kim, D.-C. \& Sanders, D. B. 1998b, ApJS, 119, 41
  \bibitem[1998]{Kim1998a} Kim, D.-C., Veilleux, S., \& Sanders, D. B. 1998a, ApJ, 508, 627
  \bibitem[2002]{Kim2002} Kim, D. C., Veilleux, S., \& Sanders, D. B. 2002, ApJS, 143, 277
  \bibitem[1992]{Kormendy1992} Kormendy, J., \& Sanders, D. B. 1992, ApJ, 390, L53
  \bibitem[1992]{Kurucz1992} Kurucz R.L., 1992, in Barbuy B., Renzini A., eds, Proc. IAU Symp. 149, The Stellar Populations of Galaxies. Dordrecht, Kluwer, p. 225
  \bibitem[1978]{Larson1978} Larson RB, \& Tinsley BM. 1978. ApJ. 219:46
  \bibitem[1989]{Lawrence1989} Lawrence, A., Rowan-Robinson, M., Leech, K. J., Jones, D. H. P., \& Wall, J. V. 1989, MNRAS, 240, 329
  \bibitem[2002]{Leitherer2002} Leitherer, C., Li, I.-H., Calzetti, D., \& Heckman, T. M., 2002, ApJS, 140, 303
  \bibitem[1997]{Lejeune1997} Lejeune T., Cuisinier F., Buser R., 1997, A\&AS, 125, 229
  \bibitem[1998]{Lejeune1998} Lejeune T., Cuisinier F., Buser R., 1998, A\&AS, 130, 65
  \bibitem[2007]{Letawe2007} Letawe, G., Magain, P., Courbin, F., Jablonka, P., Jahnke, K., Meylan, G., \& Wisotzki, L. 2007, MNRAS, 378, 83
  \bibitem[2003]{Lipari2003} L\'ipari, S., Terlevich, R., D\'iaz, R. J., Taniguchi, Y., Zheng, W., Tsvetanov, Z., Carranza, G., \& Dottori, H., 2003, MNRAS, 340, 289
  \bibitem[1987]{Lonsdale1987} Lonsdale-Persson, C. J., \& Helou, G. 1987, ApJ, 314, 513
  \bibitem[1999]{Lutz1999} Lutz D., Veilleux. S., \& Genzel, R. 1999, ApJ, 517, L13
  \bibitem[1994]{McLeod1994} McLeod, K. K., \& Rieke, G. H. 1994, ApJ, 431, 137
  \bibitem[1996]{Murphy1996} Murphy, T. W., et al.\ 1996, AJ, 111, 1025
  \bibitem[2007]{Nandra2007} Nandra K., Iwasawa K., 2007, MNRAS, 382, L1
  \bibitem[2008]{Nardini2008} Nardini, E, et al.\ 2008, MNRAS, 385, L130
  \bibitem[2010]{Nardini2010} Nardini, E., Risaliti, G., Watabe, Y., Salvati, M., Sani, E., 2010, MNRAS, inpress (arXiv:1003.0858)
  \bibitem[1998]{Natali1998} Natali, F., Giallongo, E., Cristiani, S., \& La Franca, F. 1998, AJ, 115, 397
  \bibitem[1989]{Osterbrock1989} Osterbrock, D. E. 1989, Astrophysics of Gaseous Nebulae and Active Galactic Nuclei (Mill Valley: University Science Books)
  \bibitem[2007]{Pellerin2007} Pellerin, A., \& Robert, C., 2007, MNRAS, 381, 228
  \bibitem[1980]{Richstone1980} Richstone, D. O., \& Schmidt, M. 1980, ApJ, 235, 361
  \bibitem[1996]{Sanders1996} Sanders, D. B., \& Mirabel, I. F. 1996, ARA\&A, 34, 749
  \bibitem[1988]{Sanders1988} Sanders, D. B., Soifer, B. T., Elias, J. H., Madore, B. F., Matthews, K.,Neugebauer, G., \& Scoville, N. Z. 1988, ApJ, 325, 74
  \bibitem[1983]{Schmidt1983} Schmidt, M., \& Green, R. F. 1983, ApJ, 269, 352
  \bibitem[2007]{Schneider2007} Schneider, D. P. et al.\ 2007, AJ, 134, 102
  \bibitem[2004]{Siebenmorgen2004} Siebenmorgen, R., Freudling, W., Kr\''ugel, E., \& Haas, M. 2004, A\&A, 421, 129
  \bibitem[2007]{Siebenmorgen2007} Siebenmorgen, R. \& Kr\''ugel, E. 2007, A\&A, 461, 445
  \bibitem[1980]{Sramek1980} Sramek, R., \& Weedman, D. 1980, ApJ, 238, 435
  \bibitem[1992]{Stocke1992} Stocke, J. S., Morris, S. L., Weymann, R. J., \& Foltz, C. B. 1992, ApJ, 396, 487
  \bibitem[2002]{Tacconi2002} Tacconi, L. J., Genzel, R., Lutz, D., Rigopoulou, D., Baker, A. J., Iserlohe, C., \& Tecza, M. 2002, ApJ, 580, 73
  \bibitem[2005]{Tadhunter2005} Tadhunter C. N., Robinson T. G., Gonz$\acute{a}$lez Delgado R. M., Wills K., Morganti R., 2005, MNRAS, 356, 480
  \bibitem[1972]{Toomre1972} Toomre A, \& Toomre J. 1972. Ap. J. 178:623
  \bibitem[2001]{Vanden} Vanden Berk, D. E. et al.\ 2001, AJ, 122, 549
  \bibitem[1999]{Veilleux1999} Veilleux, S., Kim, D.-C., Sanders, D.B., 1999, ApJ, 522, 113
  \bibitem[2002]{Veilleux2002} Veilleux, S., Kim, D.-C. \& Sanders, D. B. 2002, ApJS, 143, 315
  \bibitem[2004]{Veron2004} V\'eron-Cetty, M.-P., Joly, M., \& V\'eron, P. 2004, A\&A, 417, 515
  \bibitem[2006]{Veron2006} V\'eron-Cetty, M. \& V\'eron, P. 2006, A\&A, 455, 773
  \bibitem[2008]{Wang2008} Wang, J. L. 2008, Chinese J. Astron. Astrophys., 8, 643
  \bibitem[1998]{Wu1998} Wu, H., Zou, Z. L., Xia, X. Y., Deng, Z. G. 1998, A\&AS, 132, 181
  \bibitem[2010]{Yuan2010} Yuan, T.-T., Kewley L. J., Sanders D. B., 2010, ApJ, 709, 884
  \bibitem[2001]{Yun2001} Yun, M. S., Reddy, N. A., \& Condon, J. J. 2001, ApJ, 554, 803
  \bibitem[2002]{Zheng2002} Zheng, X. Z., Xia, X. Y., Mao, S., Wu, H., \& Deng, Z. G. 2002, AJ, 124, 18
  \bibitem[1999]{Zheng1999} Zheng, Z., Wu, H., Mao, S., Xia, X.-Y., Deng, Z.-G., \& Zou, Z.-L. 1999, A\&A, 349, 735
  \bibitem[1991]{Zou1991} Zou, Z., Xia, X., Deng, Z., \& Su, H. 1991, MNRAS, 252, 593

\end{thebibliography}
\end{document}